# Electrical tunability of terahertz nonlinearity in graphene

Sergey Kovalev[1], Hassan A. Hafez[2], Klaas-Jan Tielrooij[3], Jan-Christoph Deinert[1], Igor Ilyakov[1], Nilesh Awari[1], David Alcaraz[4], Karuppasamy Soundarapandian[4], David Saleta[3], Semyon Germanskiy[1], Min Chen[1], Mohammed Bawatna[1], Bertram Green[1], Frank H. L. Koppens[4], Martin Mittendorff[5], Mischa Bonn[6], Michael Gensch[7,8], and Dmitry Turchinovich[2]

[1]Helmholtz-Zentrum Dresden-Rossendorf, Bautzner Landstraße 400, 01328 Dresden, Germany

[2]Fakultät für Physik, Universität Bielefeld, Universitätsstr. 25, 33615 Bielefeld, Germany

[3]Catalan Institute of Nanoscience and Nanotechnology (ICN2), BIST & CSIC, Campus UAB, Bellaterra, 08193 Barcelona, Spain

[4]ICFO – Institut de Ciencies Fotoniques, The Barcelona Institute of Science and Technology, Barcelona, Spain

[5]Fakultät für Physik, Universität Duisburg-Essen, Lotharstraße 1, 47057 Duisburg, Germany

[6]Max-Planck-Institut für Polymerforschung, Ackermannweg 10, 55128 Mainz, Germany

[7]Institut für Optische Sensorsysteme, DLR, Rutherfordstraße 2, 12489 Berlin, Germany

[8]Institut für Optik und Atomare Physik, Technische Universität Berlin, Strasse des 17. Juni 135, 10623 Berlin, Germany

\* Email:
hafez@physik.uni-bielefeld.de;
michael.gensch@dlr.de;
dmtu@physik.uni-bielefeld.de;

**Graphene is conceivably the most nonlinear optoelectronic material. Its nonlinear optical coefficients in the terahertz (THz) frequency range surpass those of other materials by many orders of magnitude. This, in particular, allows one to use graphene for extremely efficient up-conversion of sub-THz electronic input signals into the THz frequency range at room temperature and under ambient conditions, thus paving the way for practical graphene-based ultrahigh-frequency electronic**

**technology. Here, we show that the THz nonlinearity of graphene can be efficiently controlled using electrical gating, with gating voltages as low as a few volts. For example, optimal electrical gating enhances the power conversion efficiency in THz third-harmonic generation in graphene by about two orders of magnitude. This essentially converts graphene from an almost perfectly linear, inert electronic material to a material with the highest possible THz nonlinearity. We demonstrate gating control of THz nonlinearity of graphene for both ultrashort single-cycle and quasi-monochromatic multi-cycle input signals. Our experimental results are in quantitative agreement with a physical model of graphene nonlinearity, describing the time-dependent thermodynamic balance maintained within the electronic population of graphene during interaction with ultrafast electric fields. Our results can serve as a basis for straightforward and accurate design of devices and applications for efficient electronic signal processing in graphene at ultra-high frequencies.**

## Introduction

The development of efficient broadband electronic frequency multipliers and modulators operating at the technologically important terahertz (THz) frequency range and under normal ambient conditions is of great relevance, but also challenging[1–5]. Such technologies require the integration of a highly nonlinear material in an electronic device with the possibility to control its nonlinear behavior. In addition, compatibility between the device constituents is crucial. These requirements are not easy to meet, especially in simple electronic circuits[4–6]. Here we demonstrate the possibility of accomplishing these requirements in a graphene-based device. A straightforward and efficient wide-range control of the THz nonlinearity of graphene is enabled by electrical gating of only a few volts.

One of the essential technological consequences of the Dirac-type electronic bandstructure of graphene[7,8] is its extremely nonlinear response to electric fields in the THz frequency range, which persists at room temperature and under normal ambient conditions. Graphene is a highly efficient THz nonlinear absorber, capable of displaying a nonlinear power transmission modulation of about 50% per single monolayer[9–12]. It is also an extremely efficient THz frequency multiplier, allowing for the straightforward generation of multiple THz harmonics[12–14]. Such a nonlinear response is directly attainable with quite moderate input electric fields, oscillating at frequencies not exceeding a few 100s of GHz, and with a field strength in the range of 10-100 kV/cm, which is even lower than typical channel fields in the current generation of high-speed transistors[4,15]. Conveniently, graphene is also CMOS-compatible[6,16]. This suggests hybrid graphene-CMOS technology, with the CMOS sub-THz device acting as a reliable and cost-effective pump source, providing the input signal for the efficient nonlinear processing in graphene at THz frequencies. One of the straightforward applications of such a technology is all-electronic THz frequency mixing or up-conversion of the sub-THz CMOS-generated signals into the THz range, with the potential to outcompete the current generation of diode-based ultra-high frequency mixers [1–3,17,18] in terms of conversion efficiency, operation bandwidth, compactness, and cost.

Recent experiments on THz high-harmonics generation (HHG) in graphene directly provided the measurement of its nonlinear coefficients up to the 7$^{th}$ order[14]. These nonlinear coefficients were found to exceed the respective coefficients of all other known electronic materials by many orders of magnitude[12,14]. Indeed, in a non-optimized THz HHG experiment in Ref [14], the field conversion efficiency from the driving sub-THz electric signal to THz higher harmonics of $\eta \sim 1\%$ was demonstrated in just a single atomic layer graphene sample. Such a strong THz nonlinearity of graphene is attributed to the collective thermodynamic response of its background free carrier population (electrons or holes) to the driving THz field[11,12,14]. Increasing the density of these free carriers in graphene (or equivalently, its Fermi energy $E_F$) enhances the power absorption of the THz driving field and consequently provides a control knob that allows for tuning the nonlinear response. On the other hand, too high carrier density and hence the strongly metallic phase of graphene with increased electronic heat capacity will diminish the material's thermodynamic nonlinearity. This suggests the existence of optimal doping in graphene, favoring its THz nonlinearity, which can be controlled externally. In this paper, we demonstrate such an efficient control of graphene nonlinearity by varying its free carrier density by electrical gating.

Being just one atomic layer thick and having a gapless bandstructure, graphene is a material for which wide tuning of the Fermi energy is easily achievable using a small gating voltage of only a few volts, not only in one band but also between the valence and conduction bands across the Dirac point. This offers a simple route to control the THz nonlinearity of graphene. Here we demonstrate the effects of the carrier density (or the material Fermi energy, $E_F$) on THz saturable absorption (SA) and higher-harmonics generation (HHG), as well as the associated nonlinear carrier dynamics in gated graphene. We achieve this by using nonlinear THz spectroscopy employing single- and multi-cycle THz driving fields with a peak electric field strength reaching 80 kV/cm. Within the experimentally tested range of $E_F$, increasing from ~50 to 200 meV relative to the Dirac point, transient changes in the THz field transmission of graphene in response to a THz driving field of 80 kV/cm are enhanced by a factor of ~70. For the same range of increasing $E_F$, an enhancement in the power conversion efficiency in third THz harmonic generation by almost two orders of magnitude has been achieved.

**Physical picture of gate-tunable THz nonlinearity**

The nonlinear interaction between free carriers in graphene and the field of a driving THz signal resembles a thermodynamic process exhibiting an interplay of heating-cooling dynamics[11,12,14], as illustrated in Fig. 1 for two different terahertz pulse shapes: single-cycle (Fig. 1 a) and a multi-cycle (Fig. 1 b) pulses. The THz energy absorbed in graphene by the free carriers, via intraband optical conductivity mechanism, is quasi-instantaneously (on a sub-50 fs timescale) converted into electronic heat, which leads to a rise in the carrier temperature and consequently to a suppression of both the conductivity and power absorbance (i.e. SA) of graphene [see Fig. 1(c-f)]. This heating process is then followed by a slower, picosecond-timescale carrier cooling process, occurring via emission of optical and acoustic phonons, which then restores the initial high conductivity of graphene. The interplay of the carrier heating and cooling processes during the interaction of graphene with the driving THz signal results in a temporal modulation of both the THz conductivity and the instantaneous THz absorbance [see Fig. 1(c-f)]. This leads to a nonlinear temporal deformation of the driving THz field passing through graphene, which eventually results in a spectral broadening for the single-cycle pulse, as shown in Fig. 1(g), and a highly efficient THz frequency multiplication, i.e., THz higher odd-order harmonic generation for the multi-cycle pulse, as shown in Fig. 1(h) [also see Supplement for further information].

Since the electronic THz nonlinearity of graphene is facilitated by the absorption of driving THz field by the free carriers and the subsequent thermodynamic process involving the electronic and phononic subsystems, it is the density of these free carriers that provides the key to controlling the THz nonlinearity in this material. That is, the THz nonlinearity relies on the presence of free carriers, and increasing the free carrier density increases the power absorption and leads to a stronger nonlinear response (scaling approximately with the square of the Fermi energy $E_F$), as shown in Fig. 1(b-f). However, in the regime where the density becomes extremely high (likely above $E_F = 200$ meV), the rise in the carrier temperature becomes modest, as the absorbed THz energy is to be shared among a very large number of carriers, leading to a weaker nonlinear response. Hence, there is an optimal carrier density that provides the largest nonlinearity [see Supplement for further information].

To examine the degree of control over the nonlinearity that we can expect, we first consider two levels of carrier densities corresponding to $E_F$ of 80 and 180 meV and simulate two nonlinear experiments. We use two different THz waveforms; single-cycle broadband and multi-cycle quasi-monochromatic THz pulses, as shown in Fig. 1(a & b), respectively, similar to the THz signals used in our experiments, with identical peak field strengths of 80 kV/cm. We see from Fig. 1(c-h) that for both cases of THz driving signals, the nonlinear THz field-induced effects in graphene are more pronounced for the larger $E_F$.

Another important feature presented in Fig. 1 is that for the case of multi-cycle quasi-monochromatic driving THz signal, the modulation effect in graphene is generally stronger both in amplitude and in temporal dynamics, as compared to the excitation with the single-cycle pulse, at the same value of $E_F$ and peak electric field strength in the signal. This is clearly visualized in Fig. 1(g &h) through the spectra of the THz signals transmitted through the graphene layer. The reason for the stronger nonlinear modulation of THz properties of graphene in case of multi-cycle driving field is that i) the power carried by the multi-cycle pulse is larger, assuming the same peak field as for the single-cycle, and ii) the accumulated electronic heat persists over few 100s of femtoseconds, which in turn allows for enhancing the electronic interaction with the longer-lasting multi-cycle driving signal. This presents novel opportunities for the efficient nonlinear signal processing of quasi-monochromatic THz electronic signals.

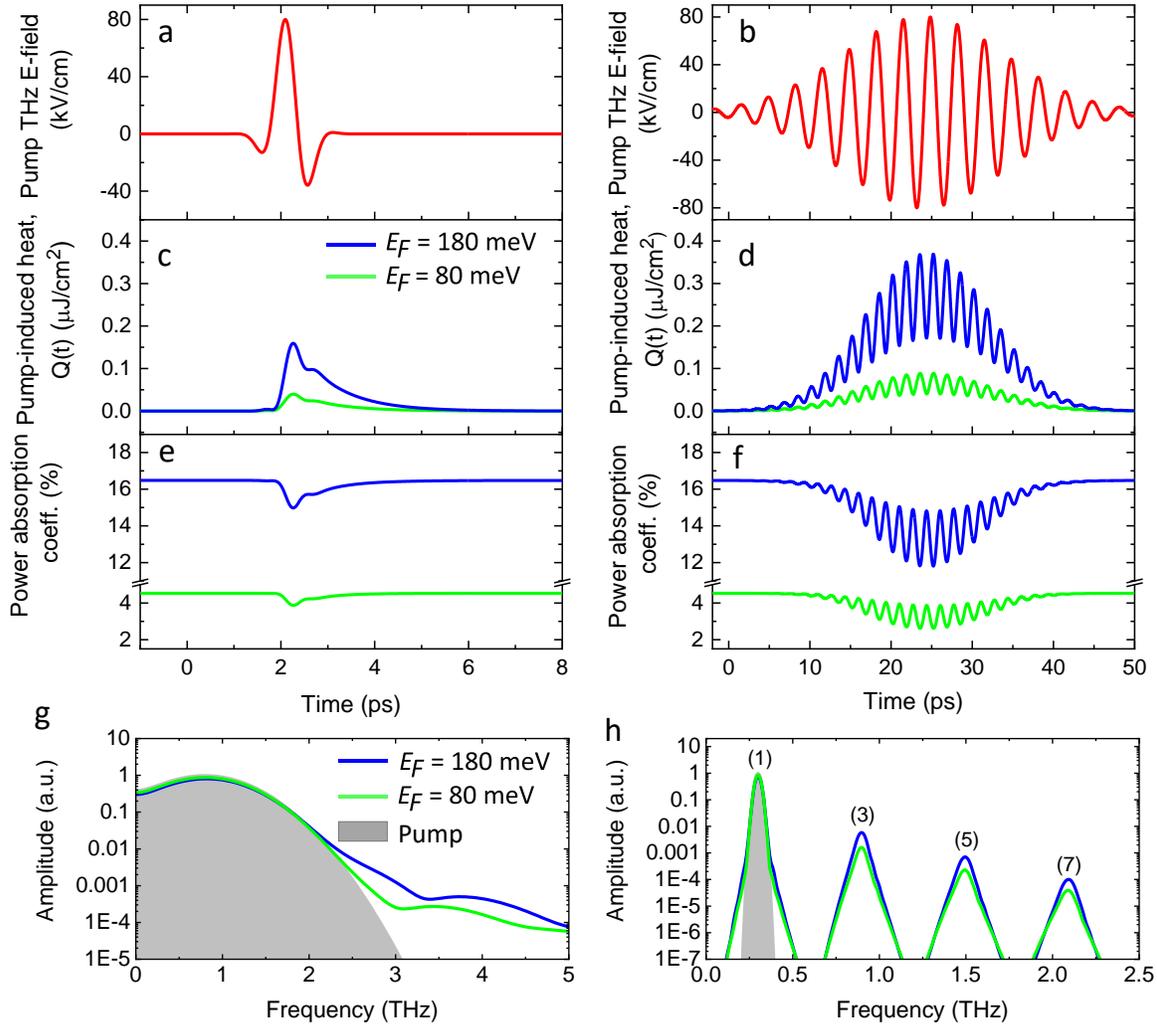

**Figure 1: Thermodynamic model calculations for the Fermi energy dependence of the THz nonlinearity of graphene. (a)** and **(b)** driving single-cycle and multicycle THz fields, respectively. **(c & d)** and **(e & f)** the associated THz field-induced heat saturable absorption, respectively, for two doping levels corresponding to Fermi energies of 80 and 180 meV. The THz induced heat (or equivalently the rise in the electron temperature) leads to a reduction and a temporal modulation of both the intraband conductivity, $\sigma(t)$ and the power absorption coefficient $\alpha(t)$, as shown in Figs. 1(c & d). This results in a nonlinear THz field-induced current $J(t) = \sigma(t, E_{THz})E_{THz}(t)$ in the graphene layer, yielding THz field-induced transparency (saturable absorption) as well as electromagnetic re-emission at higher odd-order harmonics. Both the THz induced heat and saturable absorption become more pronounced when the doping concentration (the Fermi energy $E_F$) increases, scaling nearly with $E_F^2$. **(g)** and **(h)** the spectral amplitudes of the THz fields transmitted through graphene relative to the pump field (gray background), for the tested Fermi energies of 80 and 180 meV.

Our calculations based on the thermodynamic model presented here use a single-band picture, in which only intraband free-carrier dynamics, subjected to a conservation of carrier density within the THz-excited band is considered (i.e. either electrons in the conduction band or holes in the valence band). Whereas this approach is strictly valid for graphene with a large $E_F$, it is expected to be less accurate for the situation with $E_F < k_B T_e$, where a two-band thermalization picture should instead be considered. However, we use the single-band approach universally for the whole range of tested Fermi energies, because it very well reproduces the entirety of our experimental observations using a minimum of adjustable parameters (namely, only the proportionality constant between the electron energy and electron scattering time). At the same time, the two-band model would require introduction of additional adjustable parameters (see Supplementary Information).

**Sample and nonlinear THz spectroscopy**

Based on our simulations that predict a large gate-tunability of the nonlinear response in both single-cycle and multi-cycle cases, we have conducted two types of nonlinear THz spectroscopy experiments on our electrically gated graphene sample: (i) table-top laser-based intense THz pump – THz probe (TPTP) spectroscopy using single-cycle broadband THz pulses, as shown in Fig. 2(a), and (ii) accelerator-based nonlinear THz time-domain spectroscopy (nonlinear THz-TDS) with multi-cycle quasi-monochromatic THz fields, as shown in Fig. 2(b). The former provided the measurements of the THz field-induced SA and the associated carrier (temperature) dynamics, while the latter enabled the observations of THz HHG. The graphene sample is composed of a chemical vapor deposition (CVD) – grown single-layer graphene on a 1mm-thick infrasil quartz substrate and is connected at two opposite edges to two electrodes acting as source and drain. The gating is enabled by another pair of electrodes and a polymer electrolyte (LiClO$_4$:PEO) deposited above the graphene layer and the gate electrodes, as shown in Fig. 2(c). The Fermi energy of graphene is then varied by a small gate voltage in the range from -1 to +2 V [see Methods for further information about the experiments and the sample preparation].

Figure 2(d) shows the gating response of the graphene sample, indicated by the graphene sheet electrical resistance, retrieved by measuring the source-drain current through the graphene layer at a drain voltage of 10 mV, as a function of the gating voltage. The initial doping of the graphene layer (doping at zero gate voltage) was *p*-type, as usually provided unintentionally by chemical residues during the preparation process and due to interaction with the substrate[12,19,20] , and the deposited polymer electrolyte. Here, a non-zero gate voltage $V_0$ has compensated for this hole-doping and shifted the Fermi energy to a minimum near the Dirac point, which corresponds to the peak resistance of the sample in Fig. 1(d). We note here that $E_F$ is considered relative to the Dirac point and is never uniformly zero at $|V - V_0| = 0$ due to inhomogeneity in the doping and gating field across the graphene layer, as well as a finite thermal broadening of the electron distribution, given $k_B T$ = 26 meV at room temperature $T$ = 300 K, resulting in the finite peak resistance at $|V - V_0|$ =0. Increasing the absolute gating voltage away from $V_0$ leads to an increase in the absolute value of $E_F$, whether for electrons or holes, as shown in Fig. 2(d). Quantitatively, $E_F$ is related to the free carrier density, $N_c$, through $E_F = \hbar v_F \sqrt{\pi N_c}$, where $\hbar$ is the reduced Planck's constant and $v_F = 1 \times 10^6$ m/s is the electronic Fermi velocity in graphene.

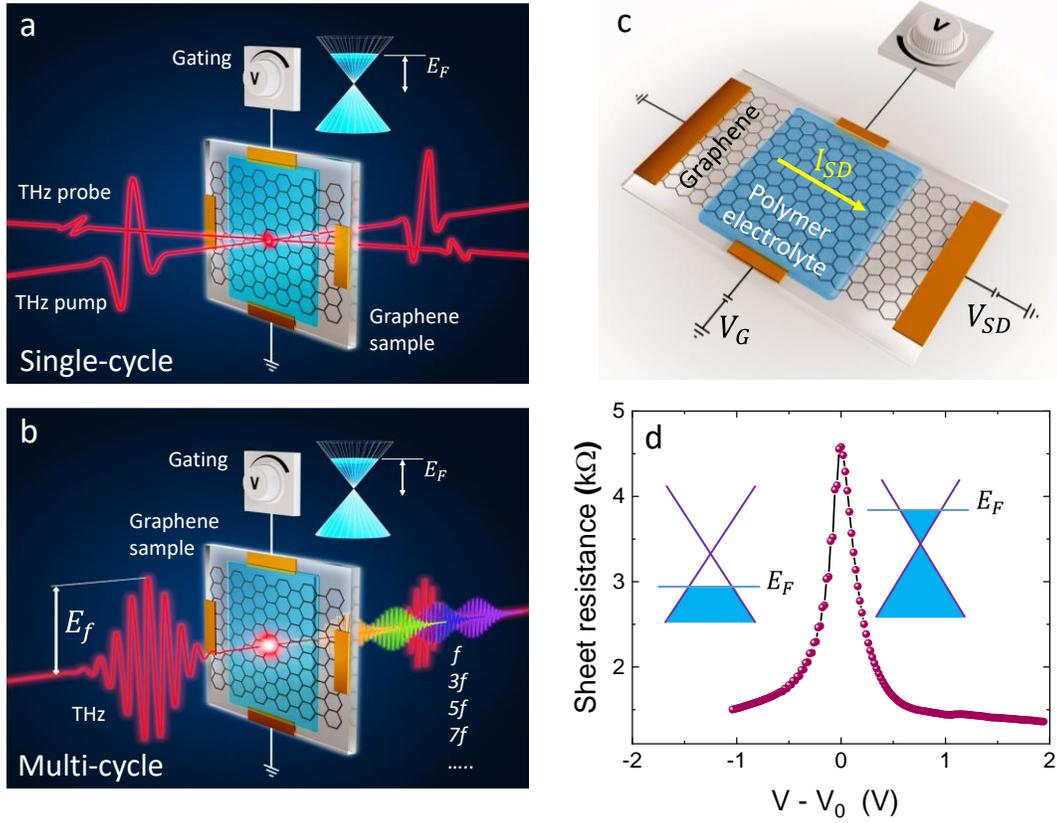

**Figure. 2: Schematics of the THz experiments and the gated graphene sample. (a)** THz-pump/THz-probe (TPTP) experiment with gated graphene. The electric field of the THz pump is vertically polarized and has a peak field strength up to 80 kV/cm, while the electric field of the THz probe is horizontally polarized with a peak field strength of less than 1 kV/cm. **(b)** Nonlinear THz-TDS, using multi-cycle quasi-monochromatic THz pulses oscillating at a fundamental central frequency of 0.3 THz and possessing a peak electric field up to 80 kV/cm. The transmitted field through the graphene sample consists of higher odd-order harmonics in addition to the fundamental frequency. **(c)** The gated graphene sample device in which the graphene film acts a channel between source and drain electrodes subjected to a constant potential difference of 0.2 mV. The graphene film is covered on top by an electrolyte subjected to a varying gating voltage to tune the Fermi level of the graphene layer. **(d)** Experimentally determined gating response of the graphene sample given by the sheet resistance of the graphene film as a function of the gating voltage relative to the voltage $V_0$ corresponding to the minimum Fermi energy in the vicinity of the Dirac point, exhibiting a maximum resistance. Positive $V - V_0$ leaves electrons in the graphene layer with the Fermi level elevated in the conduction band, while hole doping with the Fermi level pinning the valence band is induced by negative $V - V_0$.

## Results and Discussion

### Saturable absorption

Figure 3 shows both the as-measured and processed results of the single-cycle TPTP experiment, demonstrating the efficient control of the nonlinear THz transmission of graphene by using a small gating voltage of 2 Volts at maximum. Here and in the rest of this paper, we refer to $|V - V_0|$ simply as the gating voltage. The graphene sample was excited by THz pump pulses with peak electric field strength ranging from 1 to 80 kV/cm. The effects induced by the THz pump were probed by a weak THz probe pulse, at various values of $E_F$ controlled by the gating voltage. Fig. 3(a & b) show the THz probe pulses transmitted through the graphene sample with and without the intense THz pump at 80 kV/cm for two different gating voltages. After the pump, the probe transmission increases, indicating THz absorption bleaching (saturable absorption) in graphene, which corresponds to a reduction of the graphene conductivity. This effect becomes more pronounced with increasing the gating voltage (the increase in transmission of the "red" pulse relative to the "black" pulse is more pronounced in Fig. 3(b) compared to the situation in Fig. 3(a)). This is further clarified by the difference pulses shown in blue ($\Delta E = E_{red} - E_{black}$, multiplied by 4 for clarity) that show a larger pump-induced change in transmission for the case of the larger gating voltage.

Fig. 3(c) shows the as-measured transmission of the probe peak field as a function of the pump-probe delay time for various gating voltages. We note that the background probe transmission $T_0$ before the pump decreases with the gating voltage because of increasing absorption when the carrier density increases by gating. With the intense THz pump, a transient increase in the transmission occurs. Importantly, this nonlinearity becomes much more pronounced at a larger gating voltage. The secondary smaller features observed at ~16.5 ps in Fig. 3(c) are due to internal round-trip etalon reflection within the quartz substrate. Fig. 3(d) shows the analyzed THz differential transmission $\Delta T/T_0$ of the THz probe obtained from the data of Fig. 3(c), revealing that the nonlinear THz field transmission of graphene is modulated from a minimum of less than 0.1% to ~7%, i.e. by 70 times, when a gating voltage of only 2 Volts is applied. This, indeed, demonstrates an efficient, straightforward control of the graphene nonlinearity with a very modest gating voltage.

We further observe from Fig. 3(d) that the transient increase in transmission reaches a peak after ~2 ps, and is then followed by an exponential decay over a few picoseconds. The dots represent the experimental data from Fig. 3(c), while the solid lines guide the eye [see Supplement for further information]. This temporal evolution of $\Delta T/T_0$ in Fig. 3(d) is attributed to the nonlinear thermodynamic modulation of the graphene properties illustrated in Fig. 1 and its related discussion above. It is also consistent with observations reported in numerous studies where the electron temperature dynamics were ascribed to heating-cooling effects using various techniques, including time-resolved angle-resolved photoemission spectroscopy[21], TPTP[9], optical pump – THz probe spectroscopy[22–24], and time-resolved photocurrent microscopy[25,26]. We note here that the time resolution of our measurement, which is of the order of 2 ps, is limited by the noncollinear pump-probe geometry employed in the experiment, as shown in Fig. 2(a). The dependence of the peak values of $\Delta T/T_0$ on both the THz pump electric field and the gating voltage is shown in Fig. 3(e & f). The as-measured data of Fig. 3(e) were obtained, for each gating voltage, via systematic variation of the THz peak field of the pump pulses using a pair of wire-grid polarizers. We see that $\Delta T/T_0$, i.e., the nonlinear response of graphene, increases with both, the THz pump field, and the gating voltage. We further see the saturation of $\Delta T/T_0$ setting in for gating voltages exceeding 0.7 V, which mimics the behavior of the graphene sheet resistance shown in Fig. 2(d). We also note here that the data shown in Fig. 3 were collected for the negative branch of the gating voltage. On the positive branch of the gating voltage (not shown here), the sample maintains a very similar (but not perfectly the same) behavior, exhibiting a small asymmetry.

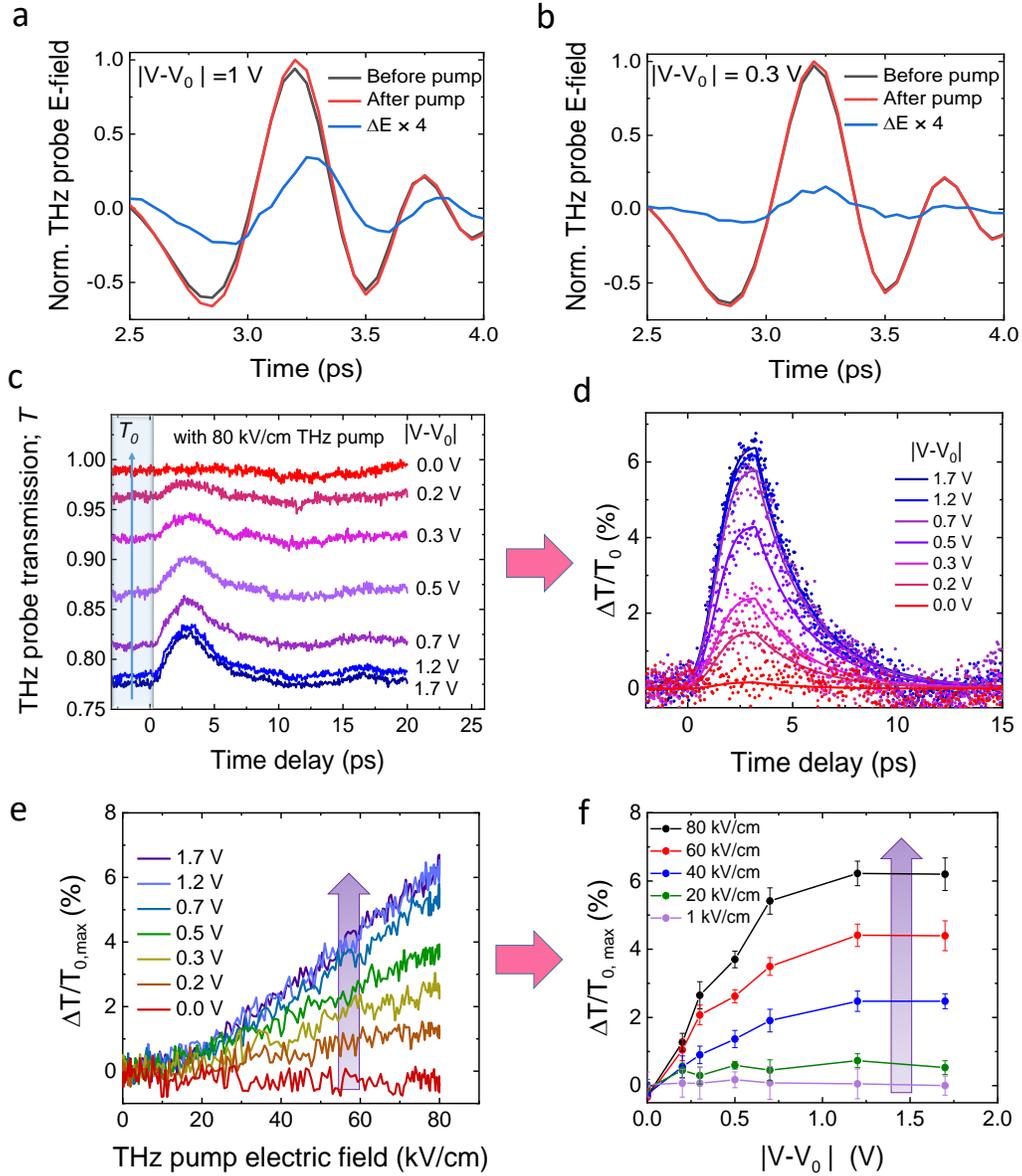

**Figure 3: Dependence of THz field-induced saturable absorption on the gating voltage revealed by TPTP spectroscopy. (a)** and **(b)** the THz probe fields before (black) and after (red) intense THz pump excitation, normalized to the peak field after excitation with 80 kV/cm pump peak electric field, and their difference (blue) multiplied by 4 for clarity, for gating voltages of 1 and 0.3 V, respectively. **(c)** The as-measured peak field transmission of the probe pulses at various gating voltages when the graphene is pumped at 80 kV/cm. $T_0$ is the background probe transmission of the graphene sample before the pump, which increases by decreasing the doping (gating) level. **(d)** the THz probe differential transmission $\Delta T/T_0 = (T-T_0)/T_0$, where $T$ is the probe transmission after the pump, as a function of the pump-probe delay time, obtained by analysis of the experimental data of (c). The dots represent the experimental results, while the solid lines are guides for the eye. **(e)** The peak of the probe differential signal $\Delta T/T_0$ as a function of the THz pump field at various doping levels, and **(f)** the peak of $\Delta T/T_0$, extracted from (e), as a function of the gating voltage at some selected pump peak electric fields. The pink arrows indicate that the data in (d) and (f) are obtained by analysis of data from (c) and (e), respectively, while the violet arrows in (e) and (f) indicate increase in $\Delta T/T_0$ with the gate voltage and the driving field, respectively.

**THz high-harmonics**

In Fig. 4, we show the results of the nonlinear multi-cycle THz-TDS experiment using the same gated graphene sample. In this experiment, the sample was excited by a multi-cycle THz field with a central frequency of 0.3 THz, as shown in the inset of Fig. 4(a). The peak electric field of the driving pulse was as high as 80 kV/cm, i.e., the same as the pump peak field in the single-cycle TPTP experiment. The response of the graphene sample was revealed through transmission measurements using calibrated free-space electro-optic sampling detection of the transmitted THz field. In Fig. 4(a), we show the frequency-domain spectra of the THz pulses transmitted through the graphene sample at various gating voltages. The spectrum includes odd-order higher harmonics of the fundamental frequency $f_1 = 0.3$ THz up to the 7th order; namely centered at the higher-harmonic frequencies $f_3 = 0.9$ THz, $f_5 = 1.5$ THz, and $f_7 = 2.1$ THz. We note here that a conservative estimate of the corresponding nonlinear electronic susceptibilities for HHG from graphene in the perturbative regime (namely at lower driving THz fields below 30 kV/cm) are as high as $\chi^{(3)} \sim 10^{-9}$ m$^2$V$^{-2}$, $\chi^{(5)} \sim 10^{-22}$ m$^4$V$^{-4}$, and $\chi^{(7)} \sim 10^{-38}$ m$^6$V$^{-6}$, for the 3rd, 5th and 7th order harmonics, respectively. These values are several orders of magnitude larger than the analogous coefficients of other known materials investigated in many different frequency ranges[12,14]. As we see from Fig. 4(a), the amplitudes of the generated harmonics, normalized to the peak at the fundamental frequency, are strongly enhanced by increasing the gating voltage from 0.14 V to 0.56 V.

Fig. 4(b) shows the dependence of the actual peak electric fields of the generated harmonics on the gating voltage, for a THz driving field of 80 kV/cm at the peak. Here, increasing the gating voltage from 0.14 to 0.56 V, leads to a magnification of the generated harmonic fields by factors of 9 for the 3rd harmonic, 4 for the 5th harmonic, and 3 for the 7th harmonic, which correspond to enhancement factors of 81, 16 and 9 for the power conversion efficiencies of these harmonics, respectively. These results display a wide range of tunability of the graphene nonlinearity and can pave ways for practical graphene-based ultrahigh-frequency electronics.

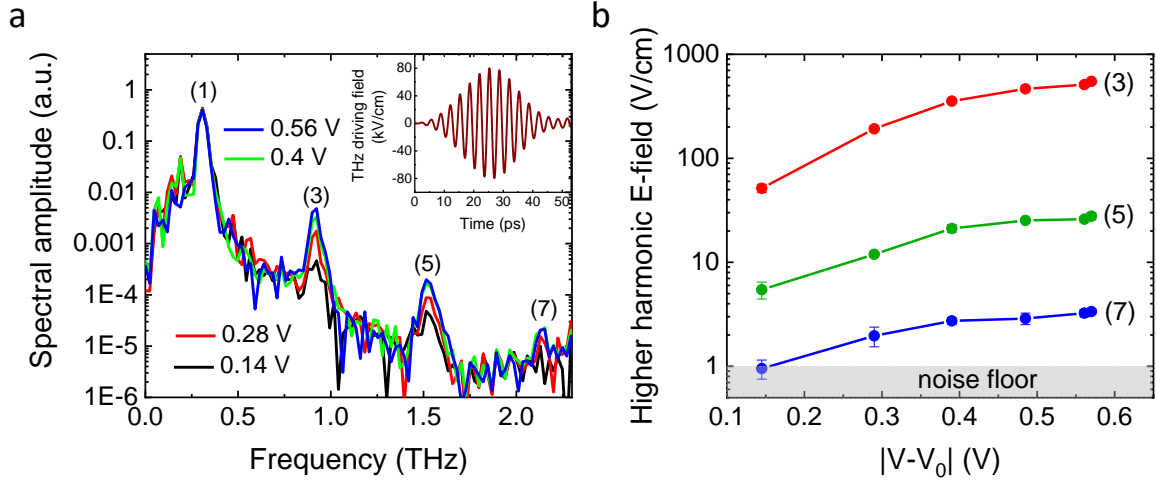

**Figure 4. Gating dependence of THz higher harmonics generation. (a)** The THz amplitude spectra of the THz fields of the incident driving field, and the transmitted fields through the graphene sample exhibiting generation of higher odd-order harmonics up to the 7th order for two doping levels, with the driving THz signal shown in the inset, and **(c)** the peak electric field of the generated harmonics as a function of the gating voltage.

**Modeling the experimental results**

We now reproduce the experimental results through simulations based on the thermodynamic model discussed above [see Supplement and Refs[12,14] for further information]. In our calculations, we simulate the nonlinear propagation of the driving THz signals, in the time domain, through the graphene layer at various Fermi energies corresponding to the experimentally tested gate voltage. In Fig. 5 (a & b), we show the model calculations (solid lines) along with the experimental results (symbols) of Figs. 3(f) and 4(c), respectively, after replacing the gating voltage $|V - V_0|$ by the corresponding $E_F$ shown in Fig. 5(c). We see a good agreement between the experiment and the model calculations for the assumed range of $E_F$ up to 200 meV, in which the graphene nonlinearity (both differential transmission and HHG) is enhanced significantly with increasing $E_F$. Above this doping level and up to $E_F \sim 260$ meV, we see from Fig. 5(a) that the experimental data shows saturation in the nonlinearity, which is also captured by the model calculations. We attribute this saturation behavior to the fact that the electronic heat capacity of graphene increases with $E_F$ at high doping levels,[27–29] restricting the increase of the carrier temperature with further increase in THz field excitation, which leads to a smaller increase in the graphene nonlinearity at such high Fermi energies [see Supplement for further information]. These effects are automatically included in the thermodynamic model by respecting the conservation of energy and total carrier density[11,12,29].

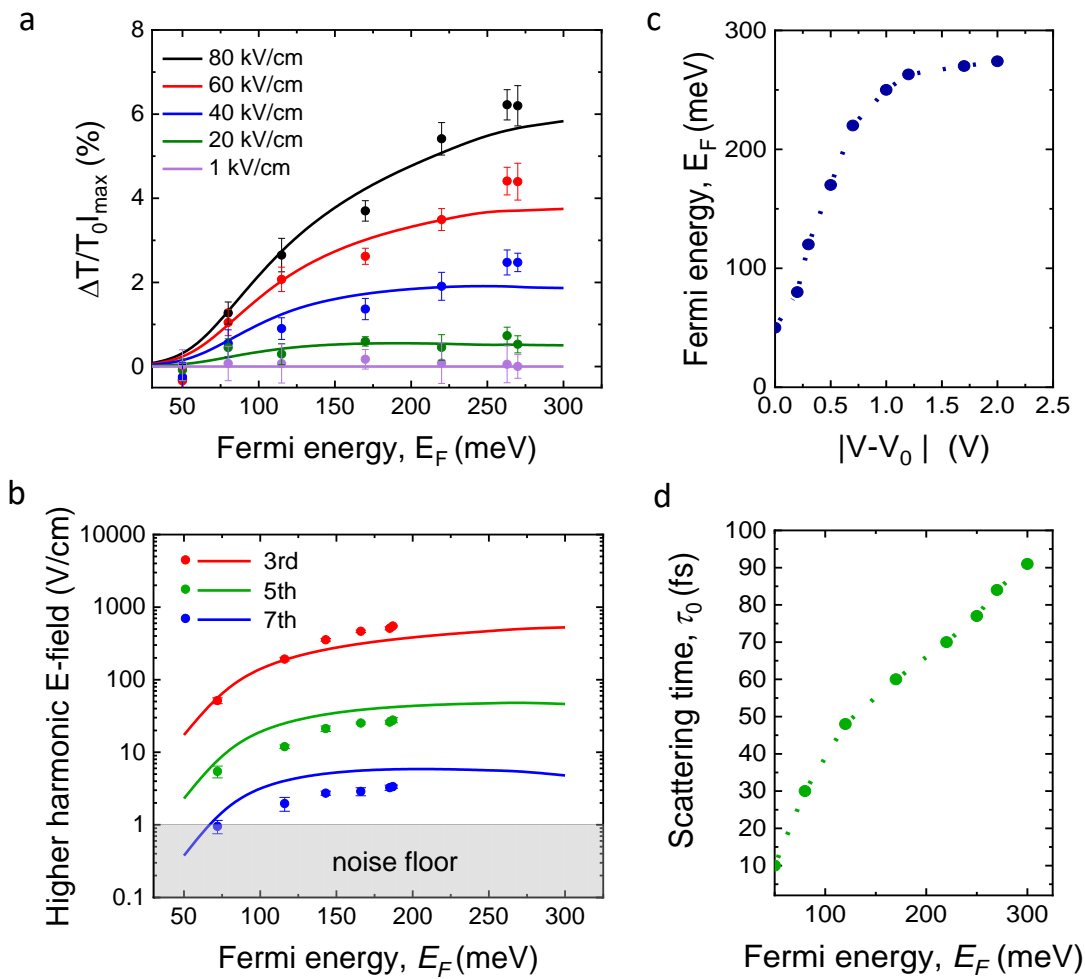

**Figure 5: The thermodynamic model calculations. (a & b)** the thermodynamic model calculations (solid lines) and the experimental results (solid circles) of the peak of $\Delta T/T_0$ and the electric field of the generated harmonics up to the seventh order, respectively, as functions of the Fermi energy and various THz pump fields, **(c)** the Fermi energy $E_F$ as a function of the gating voltage, and **(d)** the carrier momentum scattering time as a function of $E_F$.

In our model calculations, we considered a range of $E_F$ from 50 to 260 meV (as a variable parameter) to be equivalent to the values corresponding to the experimentally applied gating voltage $|V - V_0|$ in the range from 0 to 2 V, as shown in Fig. 5(c). As previously noted, $E_F$ is never uniformly zero at $|V - V_0| = 0$. Therefore, we assigned a value of $E_F = 50$ meV to the gating voltage $|V - V_0|=0$, which is consistent with the puddle energy estimates reported in previous works on similar graphene samples [24,30,31]. The initial linear increase of $E_F$ with the gating voltage is consistent with Hall measurements [see Supplement]. Another free parameter used in our calculations is the momentum scattering time $\tau_0$ associated with the linear response of graphene to low THz fields and is considered to be a function of $E_F$,[23,24,32] as shown in Fig. 5(d). Here we considered the linear THz peak transmission $T_0$ indicated in Fig. 3(c), which is related to the "frequency-integrated conductivity" of the graphene layer, to estimate $\tau_0$ and $E_F$ as carefully chosen pairs of values that reproduce the nonlinear response to the intense driving signals as well. We note that choosing other pairs of $\tau_0$ and $E_F$ in our calculations resulted in a significant deviation between the model and the experimental data. With the obtained range of $E_F$ and $\tau_0$, we extract a carrier mobility of ~3000 $cm^2V^{-1}s^{-1}$, which is consistent with values reported for typical CVD graphene[33,34]. Moreover, the increase in $\tau_0$ with $E_F$ indicates that long-range Coulomb scattering dominates the carrier scattering process[23,24,32].

**Conclusions**

The gigantic THz nonlinearity of graphene can be efficiently tuned by modest gating voltages. Our findings thus pave the way for exploiting graphene in applications that require a straightforward control of its nonlinear behavior. Therefore, applications such as graphene-CMOS hybrid devices for tunable frequency conversion, and potentially many other applications, including, for example, the development of efficient nonlinear THz modulators, shutters, switches, and conceivably other applications, are all expected to benefit from both the gigantic nonlinearity of graphene and the possibility to efficiently tune it by a small gating voltage. Interestingly, graphene also allows for electrical modulation at exceptionally high speeds, limited ultimately by the carrier cooling time of a few picoseconds, thus corresponding to a bandwidth of a few hundreds of GHz. Indeed, graphene-based amplitude[35] and phase[36] modulators with bandwidths as large as a few tens of GHz have been already demonstrated. Whereas such speeds cannot be reached with the polymer electrolyte that we employed here, it is attainable for example with a doped silicon back-gate, or by hybridizing the electrolyte-gated graphene with back-gating[37]. This opens prospects for efficient nonlinear conversion of THz signals that can be modulated with a bandwidth of tens of GHz or even higher, which is of great interest for ultrahigh-speed information and communication technologies.

## Materials and Methods

In the TPTP experiment, intense single-cycle THz pulses with a peak electric field strength up to 80 kV/cm are generated by optical rectification of 800 nm, 40 fs Ti:sapphire laser pulses in a LiNbO$_3$ crystal, following the pulse-front tilting technique[38,39], and are used as the pump pulses to excite the graphene sample. The THz probe pulses, with a much weaker peak electric field of less than 1 kV/cm, are generated by optical rectification of 800 nm pulses from the same Ti:sapphire laser in a 1-mm-thick (110) ZnTe crystal. Both pump and probe pulses are detected by free-space electro-optic sampling in another 1 mm-thick (110) ZnTe crystal. Spatial separation between the pump and probe beams after the graphene sample was attained by following a noncolinear incidence onto the sample. Combined with a cross-polarization scheme (*s*-polarization for the intense pump beam and *p*-polarization for the weak probe), this sufficiently suppressed the intense pump beam after the sample to allow for a sensitive detection of the weak probe pulses that carry the desired information about the pump-induced effects in graphene.

In the multi-cycle experiments, the graphene sample was excited by -THz pulses generated by the accelerator-based TELBE superradiant source[40,41], with a central fundamental frequency of 0.3 THz, as shown in Figs. 2(b) and 4(a). The incident and transmitted THz pulses through the graphene sample have been detected by electro-optic sampling in a 1.9 mm-thick (110) ZnTe crystal. To enable sensitive detection of the generated harmonics and reduce the dominating driving field at the detection point, a set of high-pass filters (not shown here) was used after the graphene sample. This experiment, except for the gated graphene sample examined here, is analogous to that reported in Ref[14] , and further details about this experiment are found therein.

The gate-tunable graphene sample, as shown in Fig. 1(c), was prepared as follows. We used commercial monolayer graphene (from Graphenea) prepared via chemical vapor deposition (CVD) on copper foil. Using wet transfer, we transferred a large (~1x1 cm$^2$) piece of graphene on an Infrasil quartz substrate. We then used optical lithography to define the source and drain contacts to the graphene, as well as two metal electrodes on the side, which are close to graphene without touching it. For all electrodes, we used thermal evaporation of titanium-gold. Then, we drop-casted a transparent, polymer electrolyte top-gate, consisting of polyethylene oxide (PEO) and LiClO$_4$ with 8:1 weight ratio in a solution of methanol[42] on top of the graphene and the two gate electrodes. Finally, we glued the silica substrate onto a home-built PCB with a large hole in the centre to allow for optical transmission measurements, and wire-bonded the six electrodes to six leads connected to SMA connectors. This allowed us to measure the graphene channel resistance using the electrodes in contact with graphene, while being able to electrically tune the Fermi energy by changing the voltage on the two gate electrodes [see also Supplement for Hall measurements of a similar device].

**Acknowledgments**

M.G. and B.G. acknowledge support from the European Cluster of Advanced Laser Light Sources (EUCALL) project which has received funding from the European Union's Horizon 2020 research and innovation program under grant agreement no 654220. K.J.T. acknowledges funding from the European Union's Horizon 2020 research and innovation programme under Grant Agreement No. 804349 (ERC StG CUHL), and financial support through the MAINZ Visiting Professorship. ICN2 was supported by the Severo Ochoa program from Spanish MINECO (Grant No. SEV-2017-0706). Parts of this research were carried out at ELBE at the Helmholtz-Zentrum Dresden - Rossendorf e. V., a member of the Helmholtz Association. F.H.L.K. acknowledges support from the Government of Spain (FIS2016-81044; Severo Ochoa CEX2019-000910-S), Fundació Cellex, Fundació Mir-Puig, and Generalitat de Catalunya (CERCA, AGAUR, SGR 1656). Furthermore, the research leading to these results has received funding from the European Union's Horizon 2020 under grant agreement no. 881603 (Graphene flagship Core3). We would like to thank Jochen Teichert, Ulf Lehnert and Zhe Wang for assistance. We would like to acknowledge fruitful discussions with Zoltan Mics. We thank Juan Sierra, Marius Costache, Xiaoyu Jia, and Sergio Valenzuela for assistance with the Hall measurements.

**Authors contribution**

D.T., M.G., S.K., H.A.H. and M.M. conceived the idea of the HHG experiment, while K.-J.T, D.T. and M.Bo. conceived the idea of the TPTP experiment. H.A.H. and K.-J.T. performed the experiment and analyzed the data of the TPTP spectroscopy. S.K., H.A.H., J.-C.D., I.I., N.A., S.G., M.C., M.Ba., B.G. and M.G. performed the HHG experiment. S.K., N.A. and H.A.H. analyzed the data of the HHG experiment. H.A.H., D.T. and K.-J.T. performed the modeling and main interpretation of the data. D.A., K.S., D.S., K.-J.T. and F.H.L.K. fabricated the gate-tunable graphene device and a similar device with Hall bar geometry. D.S. and K.-J.T. performed the Hall measurements. H.A.H., K.-J.T. and D.T. wrote the manuscript with contributions from M.Bo., J.-C.D., M.M. and M.G. All co-authors discussed the results and commented on the manuscript.

**Data availability -** All data needed to evaluate the conclusions in the paper are present in the paper and the Supplementary Materials.

# Supplementary Information

# Electrical tunability of terahertz nonlinearity in graphene

Sergey Kovalev[1], Hassan A. Hafez[2], Klaas-Jan Tielrooij[3], Jan-Christoph Deinert[1], Igor Ilyakov[1], Nilesh Awari[1], David Alcaraz[4], Karuppasamy Soundarapandian[4], David Saleta[3], Semyon Germanskiy[1], Min Chen[1], Mohammed Bawatna[1], Bertram Green[1], Frank H.L. Koppens[4], Martin Mittendorff[5], Mischa Bonn[6], Michael Gensch[7,8], and Dmitry Turchinovich[2]

[1]Helmholtz-Zentrum Dresden-Rossendorf, Bautzner Landstraße 400, 01328 Dresden, Germany

[2]Fakultät für Physik, Universität Bielefeld, Universitätsstr. 25, 33615 Bielefeld, Germany

[3]Catalan Institute of Nanoscience and Nanotechnology (ICN2), BIST & CSIC, Campus UAB, Bellaterra, 08193 Barcelona, Spain

[4]ICFO – Institut de Ciencies Fotoniques, The Barcelona Institute of Science and Technology, Barcelona, Spain

[5]Fakultät für Physik, Universität Duisburg-Essen, Lotharstraße 1, 47057 Duisburg, Germany

[6]Max-Planck-Institut für Polymerforschung, Ackermannweg 10, 55128 Mainz, Germany

[7]Institut für Optische Sensorsysteme, DLR, Rutherfordstraße 2, 12489 Berlin, Germany

[8]Institut für Optik und Atomare Physik, Technische Universität Berlin, Strasse des 17. Juni 135, 10623 Berlin, Germany

* Email:
hafez@physik.uni-bielefeld.de;
michael.gensch@dlr.de;
dmtu@physik.uni-bielefeld.de;

**Thermodynamic terahertz electronic nonlinearity of graphene**

The interaction of graphene with THz fields can in general be described by the solution of the Boltzmann transport equation accounting for THz field-induced intraband transitions in the Dirac-type bands of graphene. The corresponding complex-valued intraband optical conductivity is given by [11,12,32]

$$\tilde{\sigma}(\omega) = -\frac{e^2 v_F^2}{2} \int_0^\infty D(E) \frac{\tau(E)}{1 - i\omega\tau(E)} \frac{\partial f_{FD}(E,\mu,T_e)}{\partial E} dE \quad (1)$$

where $v_F = 1 \times 10^6$ m/s is the electronic Fermi velocity in graphene, $\tau(E)$ is the carrier energy-dependent scattering time, $D(E) = 2|E|/\pi(\hbar v_F)^2$ is the density of states, and $f_{FD}(E,\mu,T_e) = 1/[\exp\left(\frac{E-\mu}{k_B T_e}\right) + 1]$ is the Fermi-Dirac distribution function with chemical potential $\mu$ and carrier temperature $T_e$. We note here that the chemical potential $\mu$ is also a function of $E_F$ and $T_e$, which follows energy and carrier density conservation laws[11,12,24]. The associated power absorption coefficient, corresponding to this optical conductivity, is given by $\alpha = Re\{\tilde{\sigma}(\omega)\}/cn\varepsilon_0$, where $n$ is the real part of refractive index, $c$ is speed of light in vacuum, and $\varepsilon_0$ is vacuum permittivity. It is thus evident that both $Re\{\tilde{\sigma}(\omega)\}$ and $\alpha$ follow the same dependence on $E_F$.

In CVD-grown graphene, such as the sample used in this study, the long-range Coulomb scattering with $\tau(E) = \gamma|E|$, dominates the electron momentum scattering process[11,12,23,32], where $\gamma = \tau_0/E_F$ is a proportionality constant. As such, the solution of Eq. (1) at room temperature (with $\mu \approx E_F$) results in an approximate dependence of both $Re\{\tilde{\sigma}(\omega)\}$ and $\alpha$ on $E_F$ as $\alpha = Re\{\tilde{\sigma}(\omega)\}/cn\varepsilon_0 \approx \gamma e^2 v_F^2 v E_F^2 /cn\varepsilon_0(1 + \omega^2\gamma^2 E_F^2)$. More qualitatively, this dependence on $E_F$ can further be approximated to be quadratic, especially at the low-frequency range and low Fermi energies, as shown in Fig. S1. Hence, it is this increase in the THz absorption in graphene with increasing $E_F$ that enhances the nonlinear graphene-THz interaction.

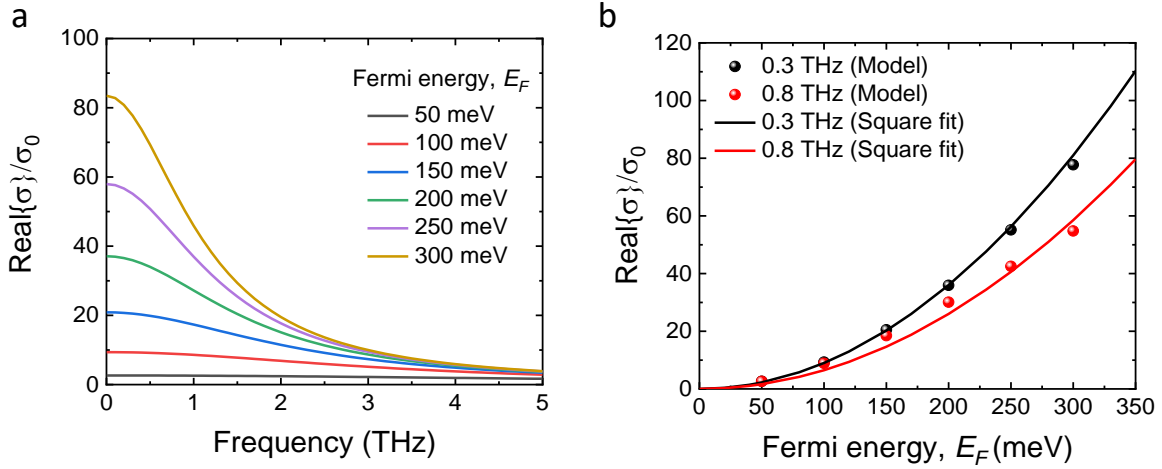

**Fig. S1:** Calculations for the Fermi energy dependence of the real conductivity of graphene. (a) the conductivity spectra normalized to the universal conductivity $\sigma_0 = e^2/4\hbar$, at various Fermi energies, calculated through Eq. (1), and (b) the normalized conductivity at two frequencies representing the central frequencies of the THz signals used in our experiment, namely 0.3 THz for the multi-cycle and 0.8 THz for the single-cycle, as a function of the Fermi energy, symbols for calculations through Eq. (1) and solid lines for $E_F^2$ dependence.

Now, with the intraband conductivity described by Eq. (1), the THz energy absorbed in graphene by free carriers results in the generation of an electrical current $J(t) = \sigma E_{THz}(t)$ in the graphene layer. Thus, the absorbed energy of incident THz signal becomes the energy of this driven current. This energy is then quasi-instantaneously converted into the electronic heat, i.e. collective kinetic energy of the electron population in graphene, via an ultrafast (sub-50 fs) thermalization process, enabled by extremely efficient electron-electron scattering in Dirac-type energy bands of graphene[21,43]. Hence, these effects become more pronounced as the field of the THz signal is increased, typically above 1 kV/cm, and thus result in a quasi-instantaneous rise in the temperature $T_e$ of the electronic system of graphene, as shown in Fig. S2(b & c). In such intraband THz conductivity dynamics in graphene, the particle number conservation rule implies that the background carrier population $N_c = \int_0^\infty D(E) f_{FD}(E, \mu, T_e) dE$ must remain constant[12], which in turn implies that the induced rise of $T_e$ leads to a simultaneous reduction both of $\sigma$ and $\alpha$. On the other hand, this quasi-instantaneous electron heating process is then followed by a subsequent electron cooling via a slower, picosecond timescale process of phonon emission, providing a channel for the recovery of the initially high conductivity and absorbance in the graphene layer. The interplay of these heating-cooling dynamics of the graphene electrons during the interaction with the driving THz signal results in a temporal modulation of the THz conductivity of graphene, i.e. $\sigma = \sigma(E_{THz}, t)$, and hence of its instantaneous THz absorbance. Thus, the corresponding driven THz current $J(t) = \sigma(E_{THz}, t)E_{THz}(t)$ becomes nonlinear with $E_{THz}$, which results in the nonlinear re-emission from the graphene layer, displaying saturable absorption and generation of odd-order higher-harmonics.

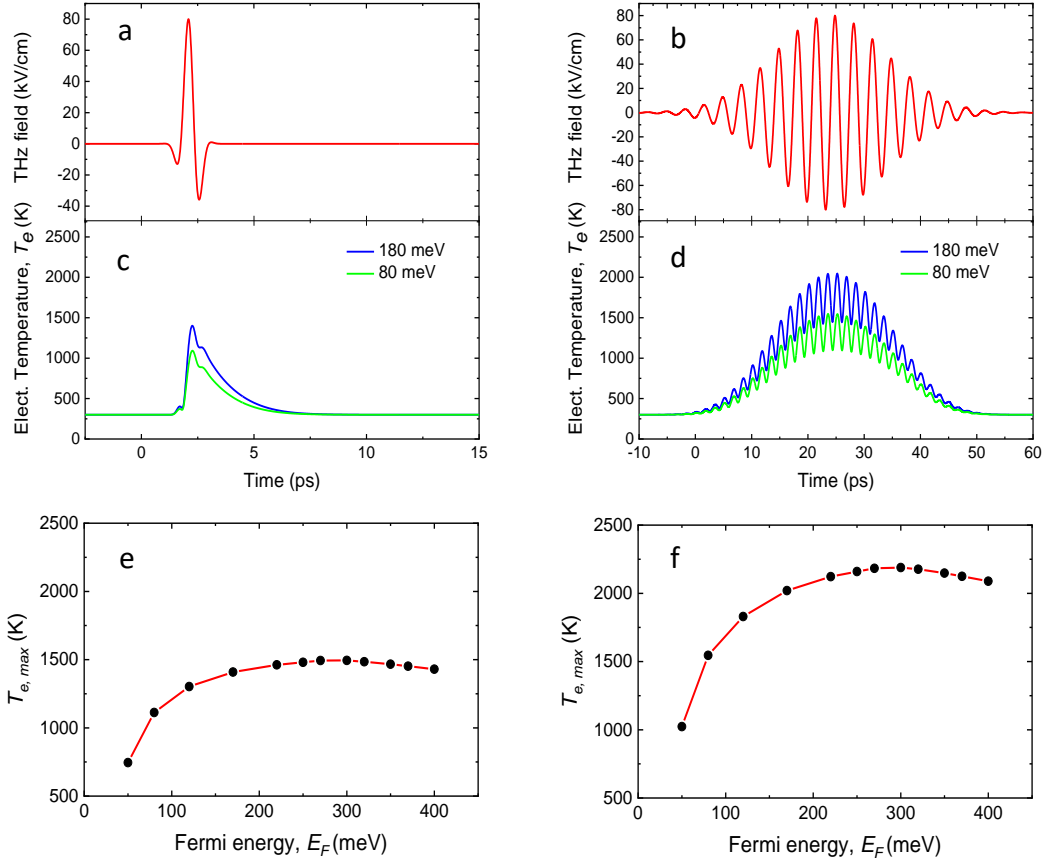

**Fig. S2:** Electronic temperature $T_e$ in graphene driven by single-cycle and multi-cycle THz signals of a peak electric field of 80 kV/cm. (a) and (b) single-cycle and multicycle THz driving signals, respectively, with a peak electric field of 80 kV/cm. (c) and (d) the electronic temperature during the propagation of the single-cycle and multi-cycle signals with graphene, respectively. (e) and (f) the dependence of the peak electronic temperature on the Fermi energy of the graphene layer.

In our calculations for the temporal evolution of these dynamics, we follow the split-step time-domain analysis described in more detail in Refs.[12,14]. In this analysis, we numerically propagate the temporal THz signal, one part after another, through the graphene film to obtain the THz field-induced heat $Q_{THz}(t)$ shown in Fig. 1(b & c), the associated electronic temperature $T_e(t)$ and the temporal modulation of the power absorbance $\alpha(t)$ and conductivity $\sigma(t)$. Fig. S2 shows the temporal evolution of the THz electronic temperature during the propagation of the single-cycle and multi-cycle THz pulses used in our experiments through the graphene layer.

The relation between the THz induced heat shown in Fig. 1(b & c) and the corresponding electronic temperature shown in Fig. S2(b-f) is given by $Q_{THz} = C_e(T_e - T_0)$, where $T_0 = 300\ K$ and $C_e = \beta T_e$ is the electronic heat capacity of graphene with $\beta = [(2\pi E_F)/(3\hbar^2 v_F^2)]k_B^2$.[27–29] We see here that $C_e$ increases with both $T_e$ and $E_F$. This means that the THz absorbed energy in graphene at largely high $E_F$ results in a smaller rise of $T_e$, as shown in Fig. S2(e & f). This explains the saturation effects in the nonlinear response of graphene shown in Figs. 3, 4 and 5 for the range of Fermi energies tested in our experiments and further predicts a reduction in the induced nonlinearity at higher Fermi energies, demonstrating the capability of the thermodynamic model to describe the THz nonlinearity of graphene.

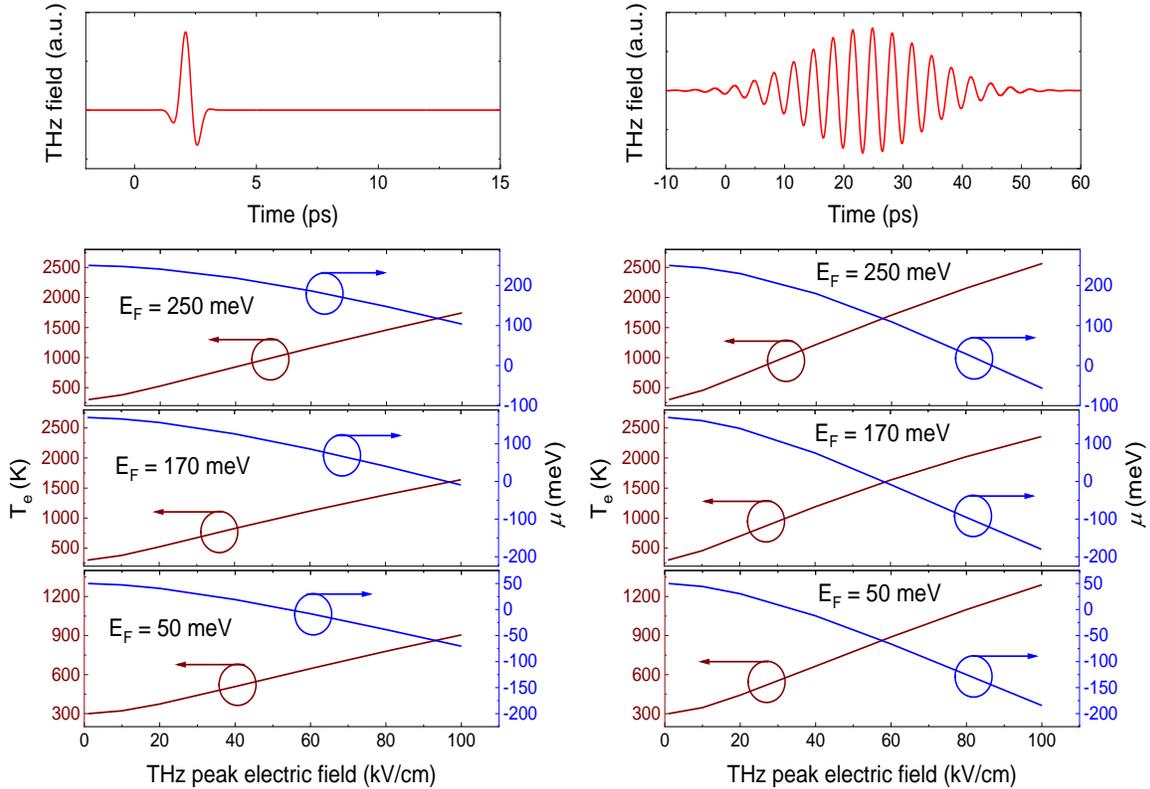

**Fig. S3:** THz peak electric field dependencies of the electron temperature $T_e$ and the chemical potential $\mu$ at various initial Fermi energies of graphene, for the single-cycle THz driving signal (the left-side panels) and the multi-cycle THz driving signal (the right-side panels).

In Fig. S3, we show the THz peak electric field dependencies of the electron temperature and the chemical potential at various initial Fermi energies, both for the single-cycle and the multi-cycle THz driving signals. The reduction in the chemical potential when the electron temperature increases fulfils the conservation of the electron density discussed above.

**Single-band versus two-band calculations**

In the calculations discussed above, we considered electron dynamics subjected to a conserved electron density in the conduction band (i.e. a single-band picture). This picture is valid in principle for $E_F > k_B T_e$. At elevated electron temperature when $k_B T_e > E_F$, the electron distribution smears out and can open a channel for electron interband transitions from the valence band to the conduction band. In this case, one may consider a two-band picture of carrier dynamics.

In Fig. S4, we show a comparison between the single-band and the two-band model calculations for a range of the electron temperature $T_e$ relevant for the thermodynamic interaction of graphene with intense THz fields. The single-band model focuses only on electron-dynamic in the conduction band, with conservation of the electron density given by $N_c = \int_0^\infty D(E) f_{FD}(E, \mu, T_e) dE = constant$, where $E = 0$ corresponds to the Dirac point. On the other hand, the two-band model is based on conserving the total electron density in the valence and conduction bands together, i.e. $N_c = \int_{-\infty}^\infty D(E) f_{FD}(E, \mu, T_e) dE = constant$. The rise in the electron temperature in both cases leads to a smearing-out of the carrier distribution that must be subjected to a reduction in the chemical potential $\mu$ in order to fulfil the aforementioned conservation rules [Fig. S4(a-c)]. In the low-temperature regime, here for $k_B T_e < 0.4 E_F$, the results for $\mu$ obtained through the two models converge and further match with the Sommerfeld approximation $\mu = E_F - \frac{\pi^2}{6} \frac{(k_B T_e)^2}{E_F}$. At higher temperatures, however, we see a disagreement between the two models, which becomes more pronounced with increasing $T_e$. Moreover, the single-band calculations show that $\mu$ further changes sign from positive to negative at high $T_e$.

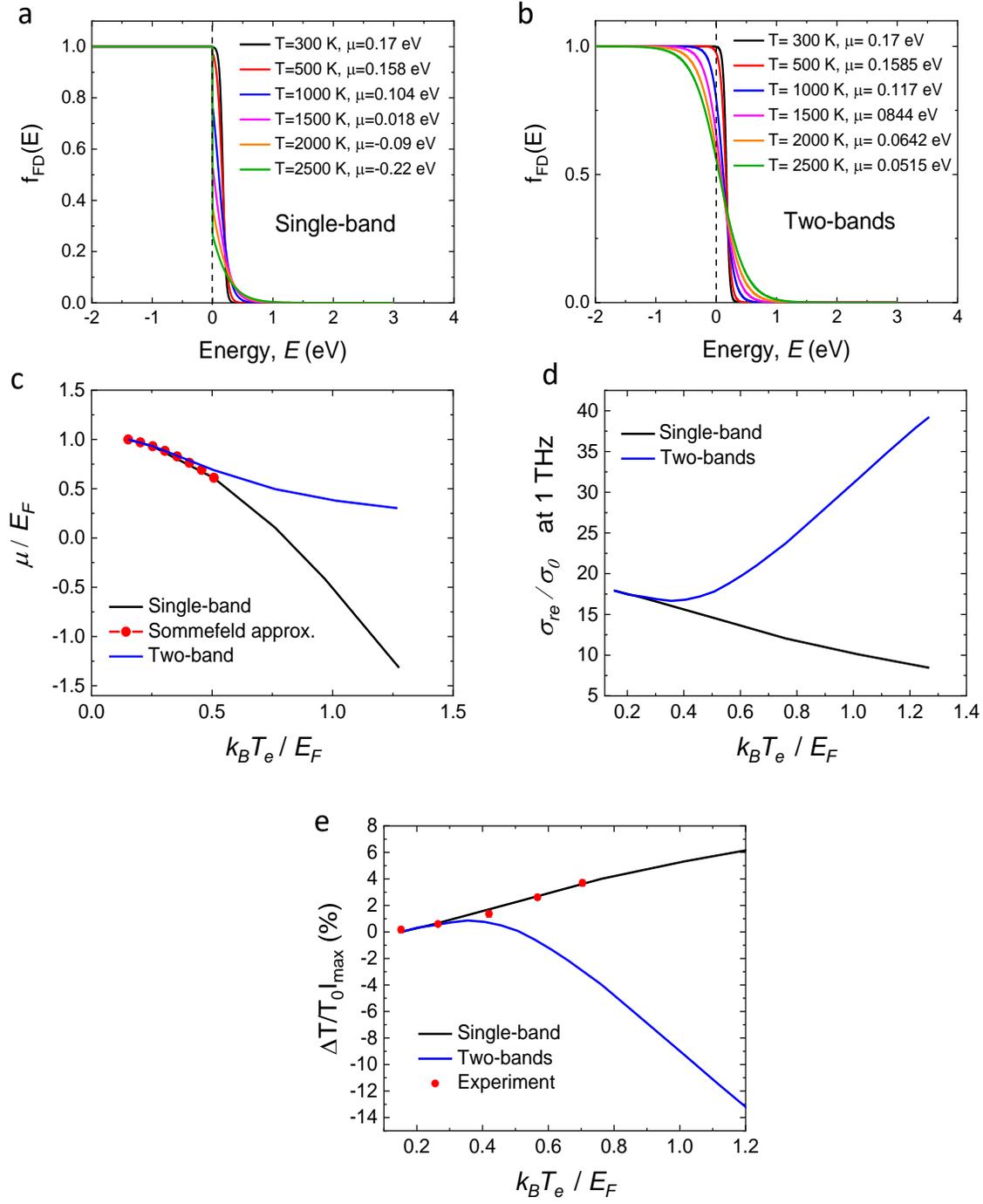

**Fig. S4:** Comparison between the single-band and the two-band calculations of the thermodynamic response of graphene to intense THz field, considering a graphene layer with a fermi energy of $E_F = 170\ meV$ and an energy dependent scattering time given by $\tau(E) = \gamma|E|$. (a) and (b) the Fermi-Dirac distribution function at various electron temperatures in view of the single-band and the two-band models, respectively, (c) the electron temperature dependence of the chemical potential, (d) the graphene conductivity at 1 THz, normalized to the universal conductivity, and (e) the peak differential transmission.

Importantly, the single-band calculations show a continuous reduction in the graphene conductivity with increasing $T_e$ [the black curve in Fig. S4(d)], which matches the experimental observations, as shown in Fig. S4(e) for the differential THz probe transmission $\Delta T/T_0$. Here, the modelled $\Delta T/T_0$ is calculated from the change in conductivity at 1 THz where the spectral amplitude of the THz probe field is a maximum. On the other hand, the two-band calculations show a small reduction in the conductivity with increasing $T_e$ for the regime with $k_B T_e < 0.4 E_F$, followed by a trend reversal displaying an increase in conductivity at higher temperatures [the blue curve in Fig. S4(d)]. This latter aspect contradicts the generally observed phenomenon of decreased THz conductivity in graphene at high THz-field excitations (or equivalently at high $T_e$) [see Refs.[9–12]], which is also observed in the present work.

This discrepancy has been addressed in several studies where the two-band model was used, for example, to explain the semiconductor-to-metal transition with positive-to-negative THz photoconductivity in photoexcited gated-graphene.[24,28] There, the momentum scattering rate was treated by considering several additional parameters and scattering contributions to reach a good agreement with the experimental observations. Thus, if we would apply the two-band model, we will need to introduce additional fitting parameters, hence making the model less rigorous. Even though it might be physically meaningful to consider additional temperature-dependent scattering rates in the two-band model, this goes significantly beyond the scope of our study. Therefore, we believe that it is better to keep the number of fitting parameters as low as possible, which is the case with our single-band calculations, especially as the discrepancy arises only in the extreme cases of very low $E_F$ and very high $T_e$.

In summary, our study of the nonlinear interaction of graphene with intense THz fields shows that the experimental observations can be well reproduced by the single-band thermodynamic model in which the number of freely adapted fitting parameters is kept to a minimum (just one in the high-field calculations, which is the proportionality constant between the scattering time and the carrier energy). The agreement between the model and the experiment is surprisingly robust, even beyond the strict applicability limits of the single-band picture. On the other hand, the two-band model that is expected to be more physically accurate has shown a significant inconsistency with the experimental observations, unless additional physical processes and corresponding fitting parameters (such as different electron scattering regimes dominating at different temperatures) are introduced.

**Guide to the eye**

Due to high noise in the experimental differential transmission $\Delta T/T_0$ of Fig. 3(d), we plotted the data in symbols and provided guide-to-the-eye lines that were generated by a function consisting of an error-function with a temporal width of 1.32 ps for the transient rise in $\Delta T/T_0$ and a bi-exponential decay function for the subsequent relaxation [see Fig. S5]. The time constants of the bi-exponential function are 2 and 3 ps, that were kept constant in the performed fitting, independent of the gating voltage.

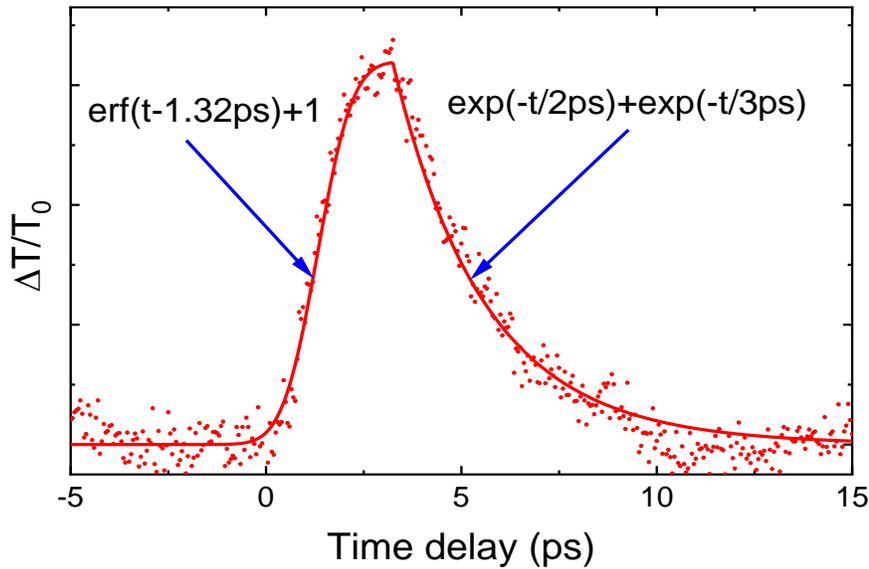

**Fig. S5:** Guide to the eye fitting with an error-function with a temporal width of 1.32 ps for the transient rise in $\Delta T/T_0$ and a bi-exponential function for the subsequent relaxation with time constants of 2 and 3 ps. The dots represent the noisy experimental data.

**Fitting parameters and supportive Hall effect measurements**

In order to test the validity of our approach to theoretically reproduce the experimental data, as shown in Fig. 5, we examine here in more detail the obtained values for the two adjustable parameters – Fermi energy $E_F$ and momentum scattering time $\tau_0$ as a function of gate voltage. We see that the Fermi energy increases roughly linearly with gate voltage, up to ~1V from the Dirac point, after which saturation occurs. We attribute the observed saturation behavior to the combination of the side electrodes and the large size of our sample. Regarding the extracted momentum scattering time as a function of Fermi energy (see Fig. 4d), we note that the observed linear scaling is in excellent agreement with what is expected for graphene whose mobility is limited by long-range Coulomb scattering, as is the case in substrate-supported CVD graphene[32]. We examined the consistency of this behavior by performing Hall measurements with a very similar device as the one used for the optical THz measurements, with the difference that this device is patterned into a Hall bar with a size of ~10 microns, rather than a size of ~1 cm of the sample used in the THz experiments. The results of Hall measurements performed on this device, as shown in Fig. S6, show that the increase in Fermi energy induced by changing the gate voltage is similar for the two devices, for voltages close to the Dirac point. Furthermore, it is very similar to the dependence reported in literature[24,28]. This adds credibility to the obtained Fermi energies as a function of gate voltage for our large device and thus to our model results. This adds credibility to the obtained Fermi energies as a function of gate voltage for our large device and thus to our model results.

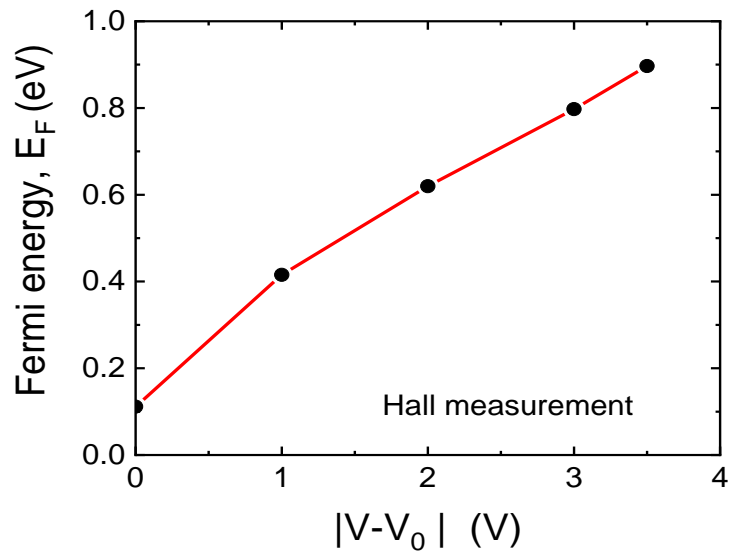

**Fig. S6:** Results of the Hall measurements; the Fermi energy as a function of the gate voltage.